\documentclass[12pt]{article}

\usepackage{graphicx}
\usepackage{amsfonts}
\usepackage{amssymb}
\usepackage{amsthm}
\usepackage{amsmath}

\usepackage[active]{srcltx}

\usepackage{here}
\usepackage{cite}

\newcommand{\ga}{\alpha}
\newcommand{\lambdat}{\tilde{\lambda}}

\begin{document}

\thispagestyle{empty}

\vspace*{3cm}
\begin{center}
{\Large{\bf
Coupling Reduction and the Higgs Mass}}

\vspace*{1cm}
{\bf N.D.Tracas$^{(1)}$, G. Tsamis$^{(1)}$, N.D. Vlachos$^{(2)}$ and G. Zoupanos$^{(1,3)}$}
\end{center}
$^1$Physics Department, National Technical University,\\
GR-157 73 Athens, Greece\vspace*{2mm}\\
$^2$Theoretical Physics Division, Aristotle University,\\
 GR-54124 Thessaloniki, Greece\vspace*{2mm}\\
$^3$Max-Planck-Institut f\"{u}r Physik,\\
 F\"{o}hringer Ring 6, 80805 M\"{u}nchen, Germany\\
and\\
Arnold-Sommerfeld-Center f\"{u}r Theoretische Physik,\\
Department f\"{u}r Physik, Ludwig-Maximilians-Universit\"{a}t M\"{u}nchen,\\
Theresienstrasse 37, 80333 M\"{u}nchen, Germany

\vspace*{1cm}
\begin{center}
{\bf Abstract}
\end{center}

\noindent
Assuming the existence of a functional relation among the Standard Model (SM) couplings gauge $\alpha_1$ and quartic $\lambda$,
we determine the mass of the Higgs particle. Similar considerations for the top and bottom Yukawa couplings in the minimal supersymmetric
SM lead to the prediction of a narrow window for $\tan\beta$,  one of the main parameters that determine the light Higgs mass.

\noindent

\vfill
\newpage
\noindent
\textit{\textbf{1. Introduction}}.
Copious theoretical efforts to establish a deeper understanding of Nature, led to very interesting constructions such as Superstring
Theories that aim to unify consistently   all interactions.  The main goal expected from a unified description of interactions
by the Particle Physics community is to understand the present day large number of free parameters of the Standard Model (SM) in
terms of a few fundamental ones. Realistically, one expects to achieve at  least a partial \textit{reduction of coupling}s.
Indeed, the celebrated SM had so far outstanding successes in all its confrontations with experimental results.
However, its apparent success is spoiled by the presence of a plethora of free parameters
mostly related to the ad-hoc introduction of the Higgs and Yukawa sectors in the theory.

Towards reducing the independent parameters of a theory, a method has been developed which looks for
renormalization group invariant (RGI) relations
\cite{KaMZ,KuMZ,KuMTZ,KMOZ,Z,Ma,Yang,Nandi,PLS,KSZ}
holding below the Planck scale, which in their turn are preserved
down to Grand Unified (GUT) or lower scales. This program applied to dimensionless couplings of supersymmetric GUTs,
such as gauge and Yukawa couplings, had already noticeable successes by predicting correctly, among other things, the top quark mass in
the finite and in the minimal N = 1 supersymmetric SU(5) GUTs \cite{KaMZ,KuMZ}.
An interesting prediction of the lightest Higgs mass in a N=1 Finite SU(5) GUT \cite{KaMZ} will be confronted with the experiment soon.
An impressive aspect of the RGI relations is that one can
guarantee their validity to all-orders in perturbation theory by studying the uniqueness of the resulting relations at one-loop,
as was proven in the early days of the  couplings reduction  program
\cite{Z}.
Even more remarkable is the fact that it is
possible to find RGI relations among couplings that guarantee finiteness to all-orders in perturbation theory
\cite{PLS}(see also \cite{EKT}).
Here, we would like to examine to which extent the above method can be applied to minimal schemes such as the SM and its minimal
supersymmetric extension, the MSSM. In fact, the former, was one of the first applications of the above reduction scheme \cite{Ma,Nandi,KSZ} assuming
a perturbative ansatz.
The implications of a stronger condition have been examined in ref \cite{Ilio}.

Let us first recall  some basic issues concerning the reduction of couplings.
A RGI relation  $\Phi(g_1,..., g_N) = 0$, has to satisfy the partial
differential equation
$\mu d\Phi/d\mu=\sum_{i=1}^N \beta_i\partial\Phi/\partial g_i=0$,
where $\beta_i$ is the $\beta$-function of $g_i$. There exist $(N-1)$ independent $\Phi$'s,
and finding the complete set of these solutions is equivalent to solve the so-called reduction equations
(REs), $\beta_g(dg_i/dg)=\beta_1,\quad i = 1,...,N$, where $g$ and $\beta_g$ are the primary coupling
and its $\beta$-function correspondingly. Using all the $(N-1)\Phi$'s to impose RGI relations, one
can, in principle, express all the couplings in terms of a single coupling g. The
complete reduction, which formally preserves perturbative renormalizability, can
be achieved by demanding a power series solution, where its uniqueness
can be investigated at the one-loop level. The completely
reduced theory contains only one independent coupling with the corresponding
$\beta$-function. This possibility of coupling unification is attractive, but it can be
too restrictive and hence unrealistic. To overcome this problem, one may use
fewer $\Phi$'s as RGI constraints.

After investigating specific examples, it becomes clear that the various couplings in
supersymmetric theories have easily the same asymptotic behavior. Therefore,
searching for a power series solution to the REs is justified. This is not the case
in non-supersymmetric theories.
Still in the SM $\ga_3$ and $\ga_2$ have the same behavior but one cannot be reduced in favor
of the other \cite{KSZ}. Here, we will examine in some detail the possibility to reduce the couplings
$\ga_1$ and the scalar quartic coupling $\lambda$ of the SM, which have the same asymptotic behavior too.

As  already mentioned, the method of reduction  was applied in the couplings of the SM in refs
\cite{Ma,Nandi,KSZ}.
The predictions for the Higgs boson mass in ref \cite{Ma,Nandi} and for the Higgs  and the top quark masses in ref \cite{KSZ}
did not survive confrontation with experiment. In the present work, after studying the evolution of the SM couplings under the
renormalization group flow, we look for solutions of the reduction equations following ref \cite{KaMZ,KuMZ,KuMTZ,KMOZ,Z,Nandi,PLS,KSZ}
by generalizing their perturbative ansatz. Eventually, we are led to the updated solutions of ref \cite{Ma}
and a Higgs mass prediction in a region that is currently under experimental investigation, which we do not consider as totally
conclusive  yet. If the experimental results persist as in\cite{ATLAS_CMS}
when better statistics are available, then
we will consider the SM case as an educative example and a motivation for applying our method in MSSM,
which is examined here too.\\

\noindent
\textbf{\textit{2. Studies of the behavior of the couplings under RGEs}}.
In the following, we will investigate  the behavior of the  SM and MSSM couplings under the renormalization group equations in order to
establish a possible realisation of the reduction scenario.The most promising case appears to connect the scalar quartic
coupling $\lambda$ and the U(1) gauge coupling $\ga_1$.
We expect that such a relation leads to a prediction of the Higgs mass.
Let us start with the 1-loop contributions. At this level,the RGE's for the gauge and the (top) Yukawa \footnote{%
Only the top Yukawa coupling is taken into account in the running.}
can be solved analytically.  The running of the quartic coupling is
governed by the equation
\begin{equation}
\label{eq:diflamda}
\begin{split}
\frac{d\tilde\lambda}{dt}&=\beta_\lambda=\frac{1}{2\pi}\left[
L_2\lambdat^2+\left(A_{1L}\ga_1+A_{2L}\ga_2\right)\lambdat+\right.\\
& \qquad \qquad \qquad \qquad A_{11}\ga_1^2+A_{12}\ga_1\ga_2+A_{22}\ga_2^2+
\left. H_L\ga_t\lambdat+H_2\ga_t^2\right],
\end{split}
\end{equation}
where
\[
\begin{split}
\tilde\lambda&=\frac{\lambda}{4\pi},\quad \ga_t=\frac{h_t^2}{4\pi},\quad t=\ln(E),\\
L_2&=6,\quad A_{1L}=-\frac32,\quad A_{2L}=-\frac92,\\
A_{11}&=\frac38,\quad A_{12}=\frac34,\quad A_{22}=\frac98,\quad
H_L=6,\quad H_2=-6,
\end{split}
\]
and $\ga_i,\,\,i=1,2,3$ are the gauge couplings.

\begin{figure}[!t]
\centering
\includegraphics[scale=0.5,angle=0]{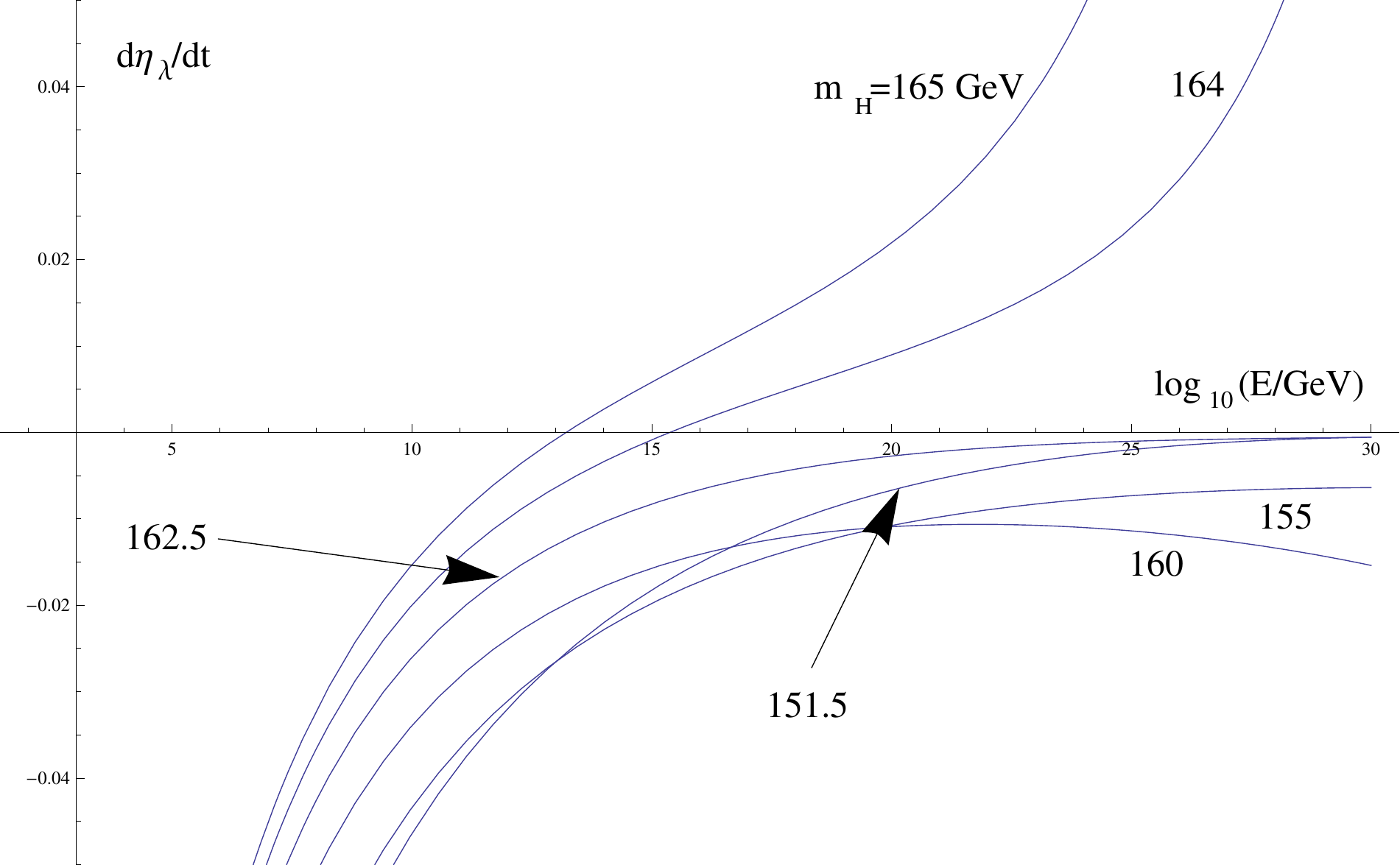}
\caption{Plotting the derivative $d\eta_\lambda/dt$ as a function of t}
\label{fig:etal-t}
\end{figure}

\begin{figure}[!t]
\centering
\includegraphics[scale=0.5,angle=0]{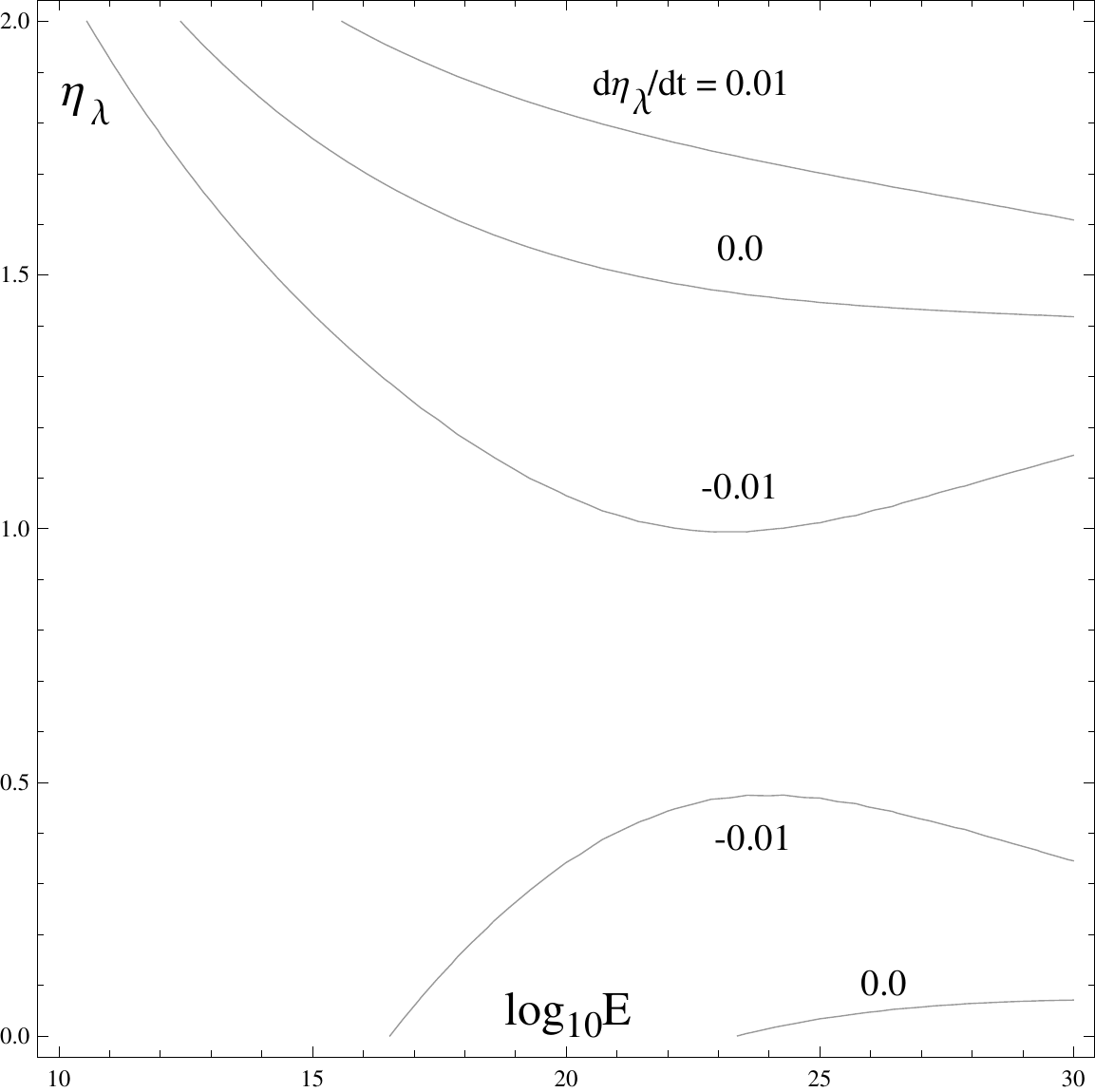}
\caption{Contours of constant value of the derivative $d\eta_\lambda/dt$ in the $(t,\eta_\lambda)$ plane
for three values: -0.01, 0 and 0.01}
\label{contours-1}
\end{figure}

To check that the ratio $\lambda$ over $\ga_1$ indeed tends to a constant value at high scales,
we plot the derivative
of the ratio $\eta_\lambda\equiv\lambdat/\ga_1$ as a function of t, for several initial values of the
$\lambdat$ coupling, which we trade for the (running) Higgs mass.
In Fig.(\ref{fig:etal-t}) we show such a plot.
Starting from $m_H=165$ GeV, we see that the derivative is positive for high energies.
Upon lowering the Higgs mass, the derivative decreases and, for $m_H\sim 162$ GeV, it
goes asymptotically to zero. Further lowering the Higgs mass the derivative becomes negative
but again for $m_H\sim 151.5$ GeV goes once more asymptotically to zero. For even smaller values
of the Higgs mass the derivative becomes positive but now $\lambdat$ passes
through negative values\footnote{%
Recall that the assumption the $\lambda$ stays always positive, for the whole energy scale, gives a lower bound
to the Higgs mass $\sim 149$ GeV.}.
Notice that the $\eta_\lambda$ becomes constant at energies well above the Planck scale, however at the 2-loop order the situation improves appreciably.
Let us explore the above situation a bit further.
We can easily express the running of the ratio $\eta_\lambda$ in the form
\begin{equation}
\label{eq:etal1}
\frac{d\eta_\lambda}{dt}=\frac{1}{\ga_1}\frac{d\lambdat}{dt}-\frac{\lambdat}{\ga_1^2}\frac{d\ga_1}{dt}=
\frac{1}{\ga_1}\beta_\lambda(\ga_1,\ga_2,\ga_t,\lambdat)-\frac{\lambdat}{\ga_1^2}\beta_1(\ga_1),
\end{equation}
where  $\beta_1$ is the 1-loop $\beta$-function for the $\ga_1$ coupling. This expression can be easily
cast in the following form
\begin{equation}
\label{eq:etal2} \frac{d\eta_\lambda}{dt}=\ga_1
\beta_\lambda(1,\ga_2/\ga_1,\ga_t/\ga_1,\eta_\lambda)-
\ga_1\eta_\lambda b_1
\end{equation}
where $\beta_1=b_1\,\ga_1^2$.
Since at the 1-loop level the differential equations for the gauge and  Yukawa
couplings  can be solved independently of the
$\lambdat$ coupling, we can express $\ga_1$, $\ga_2$ and $\ga_t$  as
functions of t and recast the above equation in the form
\begin{equation}
\label{eq:etal3}
\frac{d\eta_\lambda}{dt}=\ga_1(t)\beta_\lambda(t,\eta_\lambda)-
\ga_1(t)\eta_\lambda\beta_1(1)\equiv \ga_1(t) F_{\eta_\lambda}(t,\eta_\lambda),
\end{equation}
using the same symbol $\beta_\lambda$ for the new function of $t$
and $\eta_\lambda$.
In Fig.\ref{contours-1} we plot contours of constant value ( -0.01, 0 and 0.01) for $\ga_1(t)F_{\eta_\lambda}(t,\eta_\lambda)$
in the $(t,\eta_\lambda)$ plane. We clearly see that the zero value contour tends, for albeit very high energies, to a
constant value for the ratio $\eta_\lambda$ ($\sim 1.3$ and $\sim
0.05$).

\begin{figure}[!t]
\centering
\includegraphics[scale=0.4,angle=0]{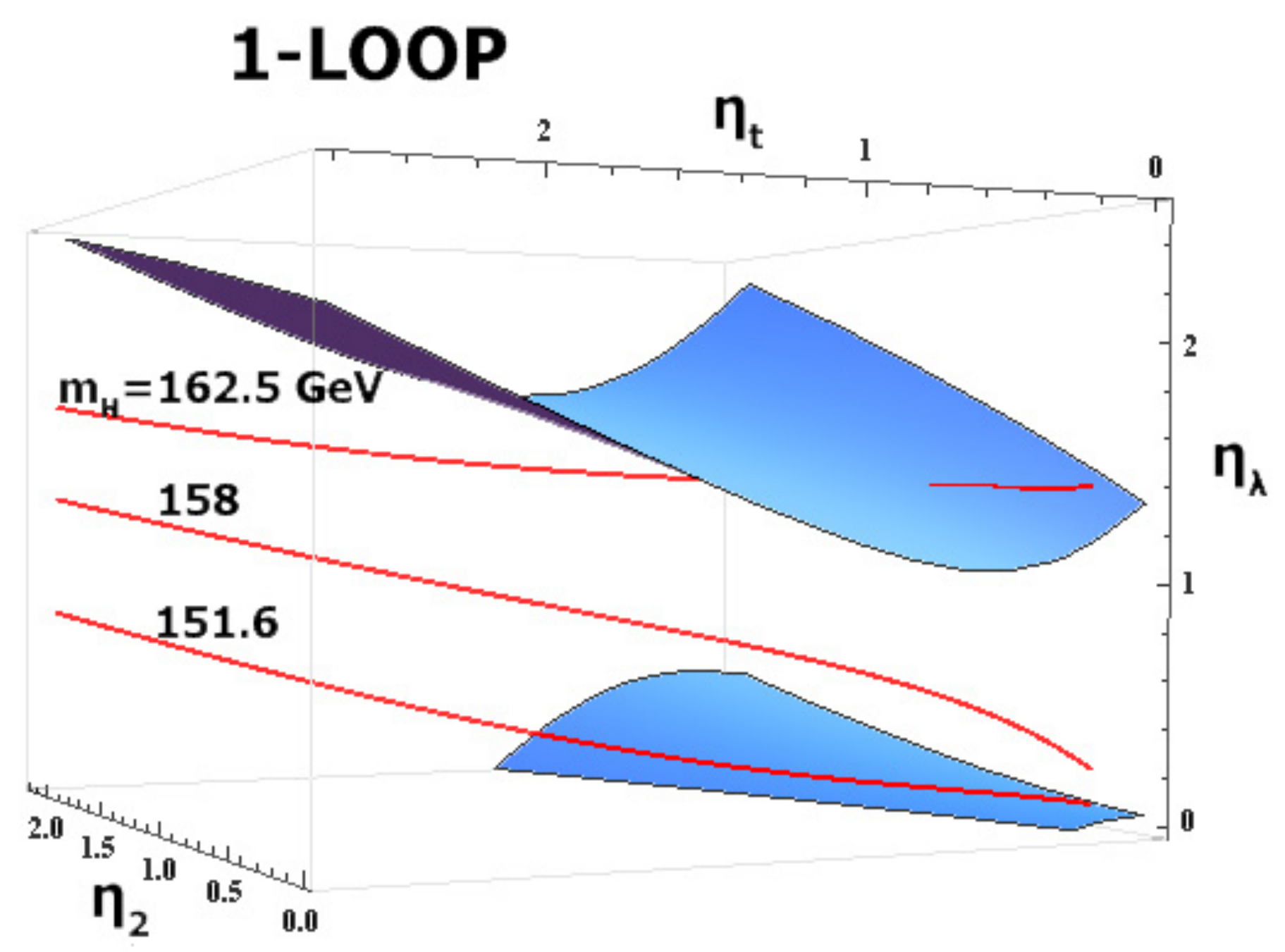}
\caption{Surfaces of constant $\eta_\lambda$ and parametric curves of $(\eta_2(t),\eta_t(t),\eta_\lambda(t))$
for three values of the Higgs mass (1-loop).}
\label{fig:surfaces-1}
\end{figure}

Let us explore this situation from even another point of view and treat $\ga_1$, $\ga_2/\ga_1\equiv\eta_2$, $\ga_t/\ga_1\equiv\eta_t$
and $\eta_\lambda$ as independent variables.
Then we rewrite Eq.\ref{eq:etal2} in the form
\begin{equation}
\label{eq:etala1}
\frac{d\eta_\lambda}{dt}=\ga_1\,F_{\eta_\lambda}(\eta_2,\eta_t,\eta_\lambda)
\end{equation}
using again the same symbol $F_{\eta_\lambda}$. The derivative of $\eta_\lambda$ with respect to $\ga_1$ is given by
\begin{equation}
\label{eq:etala2}
\frac{d\eta_\lambda}{d\ga_1}=\frac{\frac{d\eta_\lambda}{dt}}{\frac{d\ga_1}{dt}}=\frac{\ga_1\,F_{\eta_\lambda}(\eta_2,\eta_t,\eta_\lambda)}{b_1\ga_1^2}=
\frac{F_{\eta_\lambda}(\eta_2,\eta_t,\eta_\lambda)}{b_1\ga_1}.
\end{equation}
If $\eta_\lambda$ tends to a constant value, then the above derivative should tend to zero. This is of course
true when $\ga_1$ becomes very large but also when the numerator, $F_{\eta_\lambda}(\eta_2,\eta_t,\eta_\lambda)$ is equal to zero.
Just to have a first impression, we put $\eta_2=\eta_t=0$ (both ratios tend to zero for very high energies).
Then $F_{\eta_\lambda}(0,0,\eta_\lambda)$ is just a second order polynomial in $\eta_\lambda$ with zeros at $\sim 1.34$ and $\sim 0.047$,
which are the two fixed points  observed many years ago\cite{Ma}.
We can plot, in the space of $(\eta_2,\eta_t,\eta_\lambda)$, the surface where $F_{\eta_\lambda}(\eta_2,\eta_t,\eta_\lambda)=0$.
We can also  numerically solve
the differential equation and express $\eta_\lambda$ as a function of $t$. Then we can make a parametric plot of the curve
$(\eta_2(t),\eta_t(t),\eta_\lambda(t))$. We expect that for high energies, i.e. low values of $\eta_2$ and  $\eta_t$,
the curve will lie on the surface  $F_{\eta_\lambda}=0$. This is shown in Fig.\ref{fig:surfaces-1}. There are two surfaces
corresponding to $F_{\eta_\lambda}=0$ and we have plotted  the parametric curves for three Higgs masses. We clearly see that for the values
$m_H\sim 162.5$ and $151.6$ GeV, the parametric curves lie on the surfaces for low values of $\eta_2$ and $\eta_t$.\\

\begin{figure}[!b]
\centering
\includegraphics[scale=0.7,angle=0]{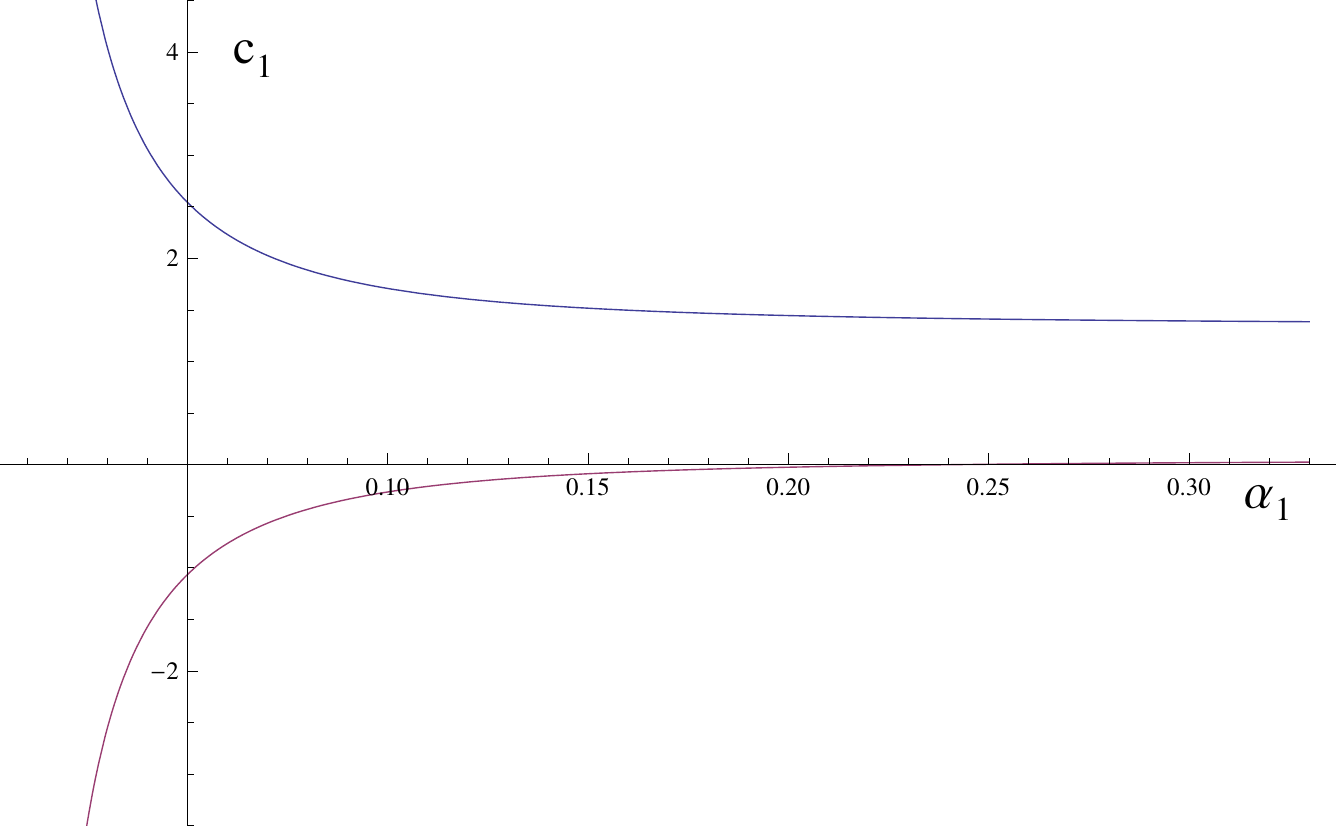}
\caption{The ``constant'' $c_1$ as a function of $\ga_1$}
\label{fig:red2}
\end{figure}

\noindent
\textbf{\textit{3. The Reduction Equations}}.
The observations made in section 2 suggest that at least the couplings $\tilde \lambda$ and $\ga_1$ are not independent in the SM
and there may exist a functional relation among them at high scales. It is therefore justified to look for solutions of
the reduction equation
\begin{equation}
\label{eq:red}
\frac{d\lambdat}{d\ga_1}=\frac{\beta_\lambda}{\beta_1}.
\end{equation}
Let us first look for solutions of Eq. 7 at 1-loop
\begin{equation}
\label{eq:ansatz1}
\lambda(t)=c_1 \ga_1(t),
\end{equation}
where $c_1$ would be a constant in the perturbative ansatz of ref \cite{KaMZ,KuMZ,KuMTZ,KMOZ,Z,Nandi,PLS,KSZ},
but here we are searching for more general solutions.
 From the 1-loop $\ga_1(t)$
we can solve for $t$ and express $\ga_2(t)$ and $\ga_t(t)$ (which
are present in $\beta_\lambda$) as functions of $\ga_1$. Using the
ansatz given in Eq.\ref{eq:ansatz1}, Eq.\ref{eq:red} becomes a second order polynomial in
$c_1$, where, of course, the coefficients depend on $\ga_1$. In
Fig.\ref{fig:red2} we plot the two solutions of the polynomial as a
function of $\ga_1$. We clearly see that for large values of $\ga_1$
(i.e. large energies), the two solutions tend to constant values.
This is easily understood, since for high energies, we can neglect
all the couplings but $\ga_1$ itself, and Eq.\ref{eq:red} reduces to
\begin{equation}
\label{eq:red2}
c_1=\frac{L_2 \ga_1^2 c_1^2 + c_1 A_{1L} \ga_1^2 + A_{11}\ga_1^2}{b_1 \ga_1^2}=
\frac{L_2  c_1^2 + c_1 A_{1L}  + A_{11}}{b_1}
\end{equation}
with the two solutions being independent of $\ga_1$ (1.34233 and 0.0465609).
We have already encountered this behavior when examining Eq.\ref{eq:etala2}. The second order polynomial above is
the $F_{\eta_\lambda}(0,0,\eta_\lambda)$.
It is worth noting that the values of $\ga_1$, when $c_1$ approaches one of
its fixed points correspond to energies well above the Planck scale. (At the Planck scale $\ga_1 \sim 0.017$ !).

\begin{figure}[!t]
\centering
\includegraphics[scale=0.5,angle=0]{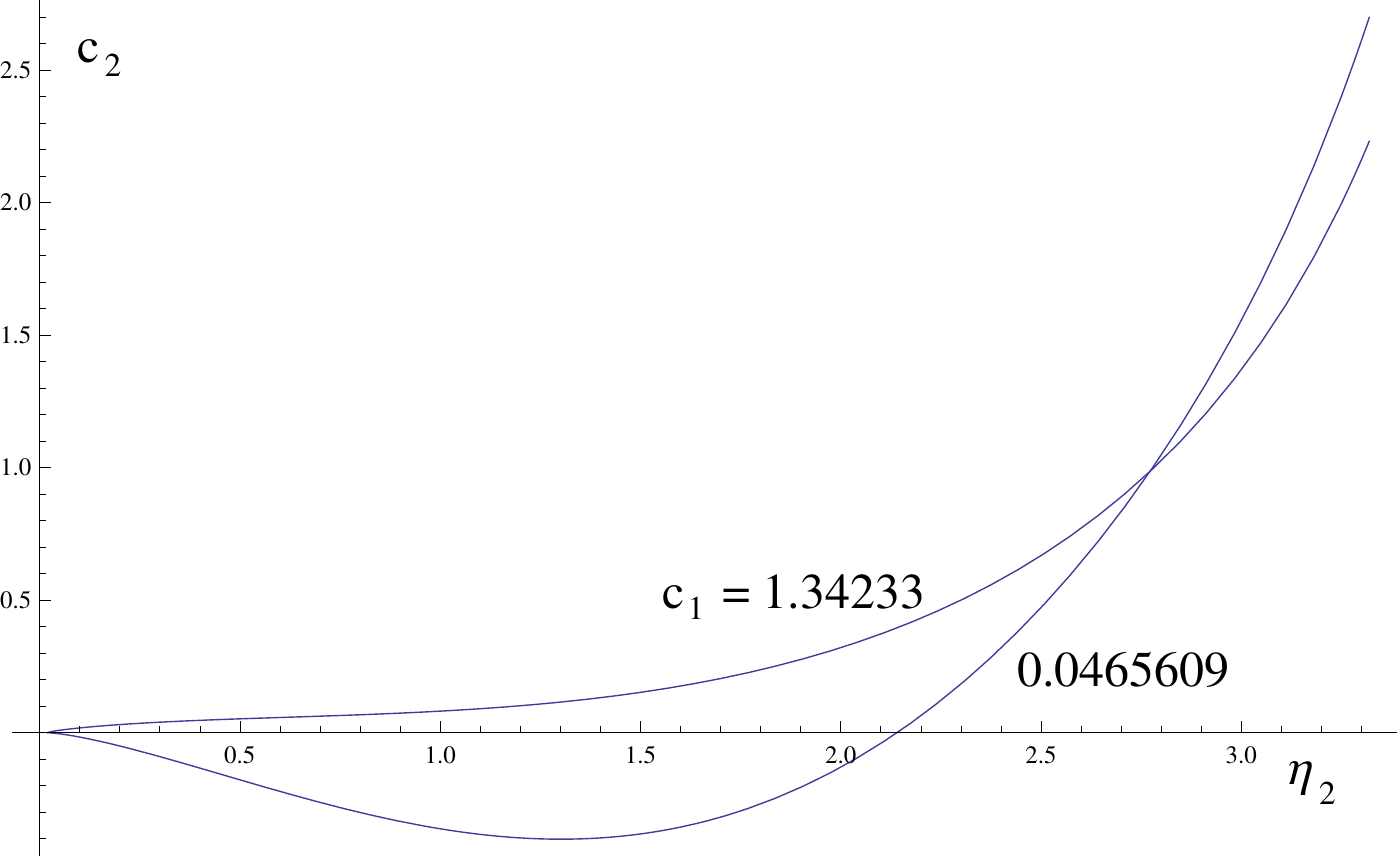}
\caption{Plotting of $c_2$ as a function of $\eta_2$ for the two values of $c_1$ (1-loop)}
\label{fig:c2_eta}
\end{figure}

\begin{figure}[!t]
\centering
\includegraphics[scale=0.5,angle=0]{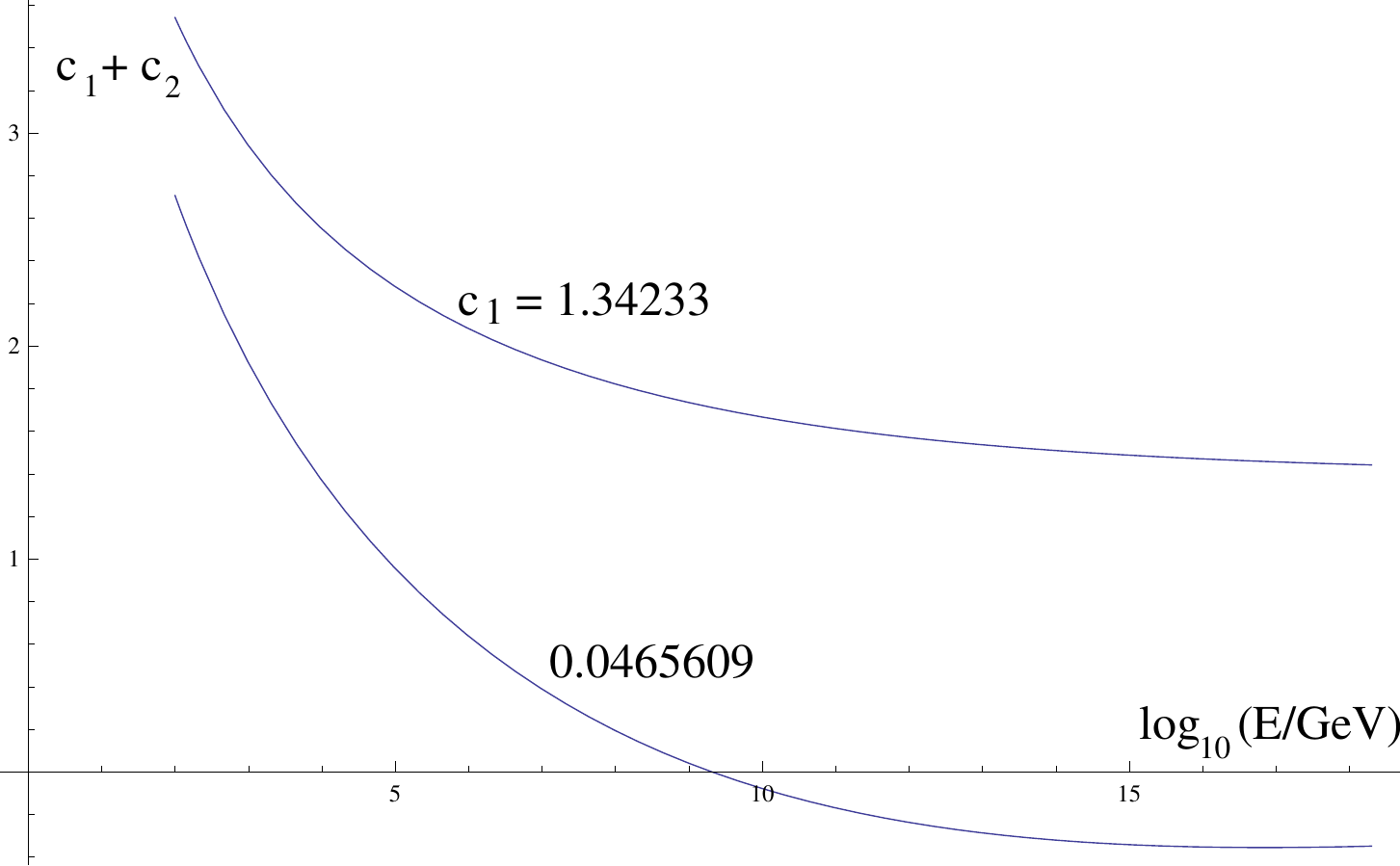}
\caption{Plotting of $c_1+c_2=\eta_\lambda$ as a function of
$\log_{10}(E)$ for the two values of $c_1$ (1-loop)}
\label{fig:c2_c1_Energy}
\end{figure}

\begin{figure}[!b]
\centering
\includegraphics[scale=0.6,angle=0]{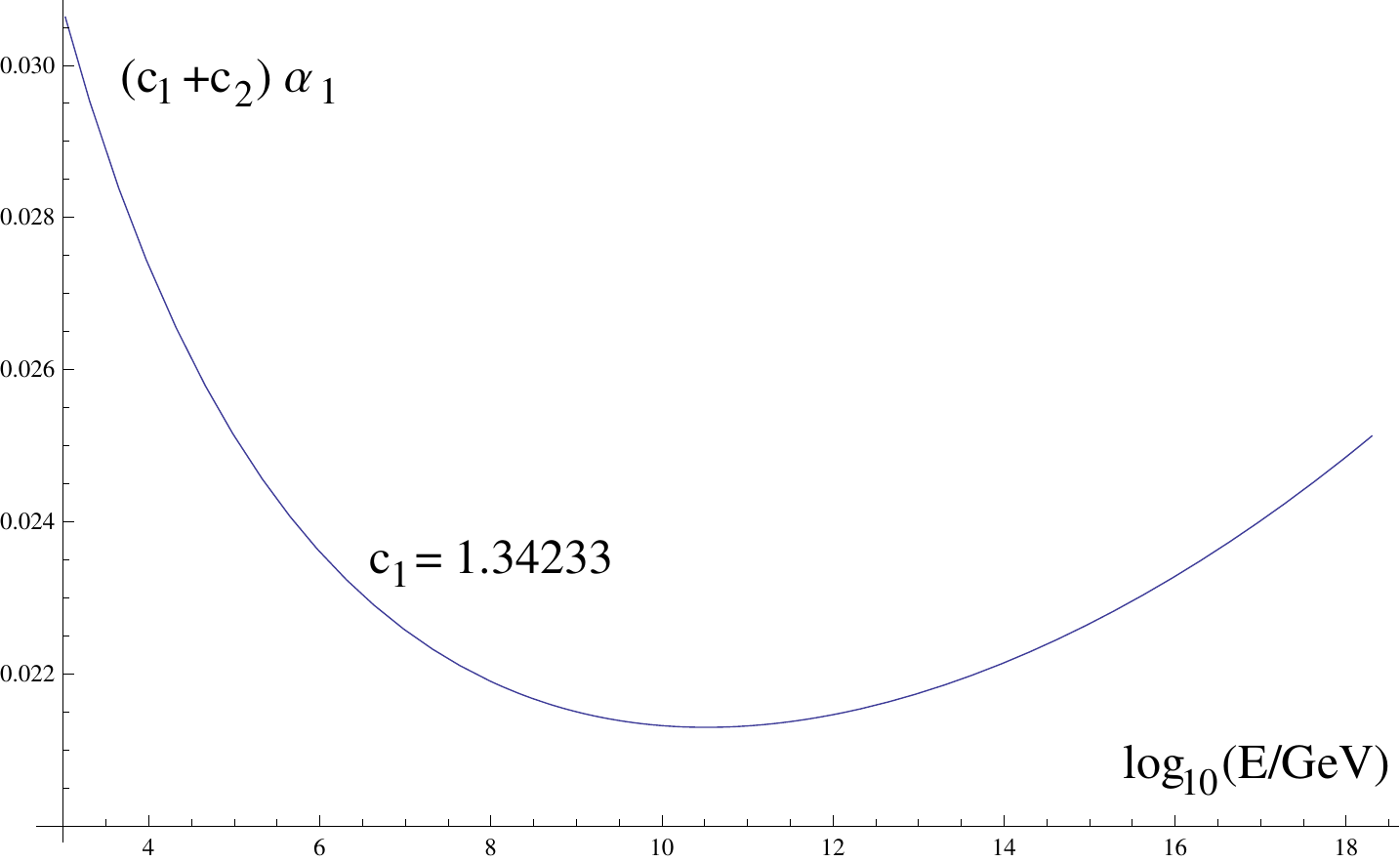}
\caption{Plotting of $(c_1+c_2)\ga_1=\lambdat$ as a function of $\log_{10}(E)$ for the higher value of $c_1$.
The corresponding running Higgs mass is $\sim 162$ GeV  (1-loop).}
\label{fig:lambda_E}
\end{figure}

We can go one step further and postulate that
\begin{equation}
\label{eq:ansatz2}
\eta_\lambda=\frac{\lambdat}{\ga_1}=c_1+c_2(\eta_2).
\end{equation}
 For high energies, the ratio $\eta_2$ tends to
zero and in order to obtain our first ansatz, we should require that $c_2(\eta_2\rightarrow 0)=0$. From Eqs.\ref{eq:ansatz2} and
\ref{eq:etal2} we easily get
\begin{equation}
\label{eq:eqtal3}
\frac{dc_2}{dt}=\frac{d\eta_\lambda}{dt}=\ga_1 \beta_\lambda(1,\eta_2,\eta_t,\eta_\lambda)-
\ga_1\eta_\lambda b_1.
\end{equation}
Writing
\begin{equation}
\label{eq:eta}
\frac{d\eta_2}{dt}=\frac{1}{\ga_1}\frac{d\ga_2}{dt}-\frac{\ga_2}{\ga_1^2}\frac{d\ga_1}{dt}=
\ga_1\left(b_2\eta_2^2-\eta_2 b_1\right),
\end{equation}
where $b_2$ is the one loop $\beta$-function coefficient for $\ga_2$, and
dividing the last two equations we get the derivative of $c_2$ with respect to $\eta_2$. All that remains to be done is to express
the ratio $\eta_t$ as a function of $\eta_2$. Having the 1-loop analytical expressions for $\ga_t$
and  $\ga_1$ as  functions of $t$, we can substitute $t$ from the relation
\begin{equation}
\label{eq:t}
\begin{split}
\eta_2&=\frac{\ga_2}{\ga_1}=\frac{\displaystyle{\frac{\ga_{20}}{1-\frac{b_2}{2\pi}\ga_{20}(t-t_0)}}}
                               {\displaystyle{\frac{\ga_{10}}{1-\frac{b_1}{2\pi}\ga_{10}(t-t_0)}}}\rightarrow\\
t&=t_0+\frac{\eta_{20}-\eta_2}{\displaystyle{\frac{1}{2\pi}\left[\eta_0 b_1\ga_{10}-\eta_2 b_2 \ga_{20}\right]}},
\end{split}
\end{equation}
where $\eta_{20}=\ga_{20}/\ga_{10}$ and $\ga_{10}$ and $\ga_{20}$ are the corresponding values at the scale $t_0$.
Substituting  $\eta_\lambda=c_1+c_2(\eta_2)$ and solving the differential equation for $c_2(\eta_2)$, we
get $c_2(\eta_2)$. In Fig.\ref{fig:c2_eta} we show the solutions for the two choices of $c_1$ using the initial condition
$c_2(\eta_2=0.2)=0$ (see Fig.\ref{fig:red2}).
In Fig.\ref{fig:c2_c1_Energy} we show $c_1+c_2$ (i.e. $\eta_\lambda$) as a function of the energy scale.
The curve which corresponds to the higher $c_1$ value has almost reached that value at the Planck scale,
while the one that corresponds to the lower $c_1$ value, apart from passing through unacceptable negative values,
is still far away from that value.
In Fig.\ref{fig:lambda_E} we plot the function $(c_1+c_2(t))\ga_1(t)$, i.e. $\lambdat(t)$ itself, for the higher $c_1$ value curve.
The corresponding running (pole\footnote{%
For the known value of the top mass and the specific region of the Higgs mass, the Higgs  pole mass is lower
than the running mass by an amount of $\sim 4.6-4.7\%$. The relation between running and pole mass can be found
in the references \cite{running-pole}.
}
) Higgs mass  is $\sim 162(154)$ GeV.

\begin{figure}[!t]
\centering
\includegraphics[scale=0.5,angle=0]{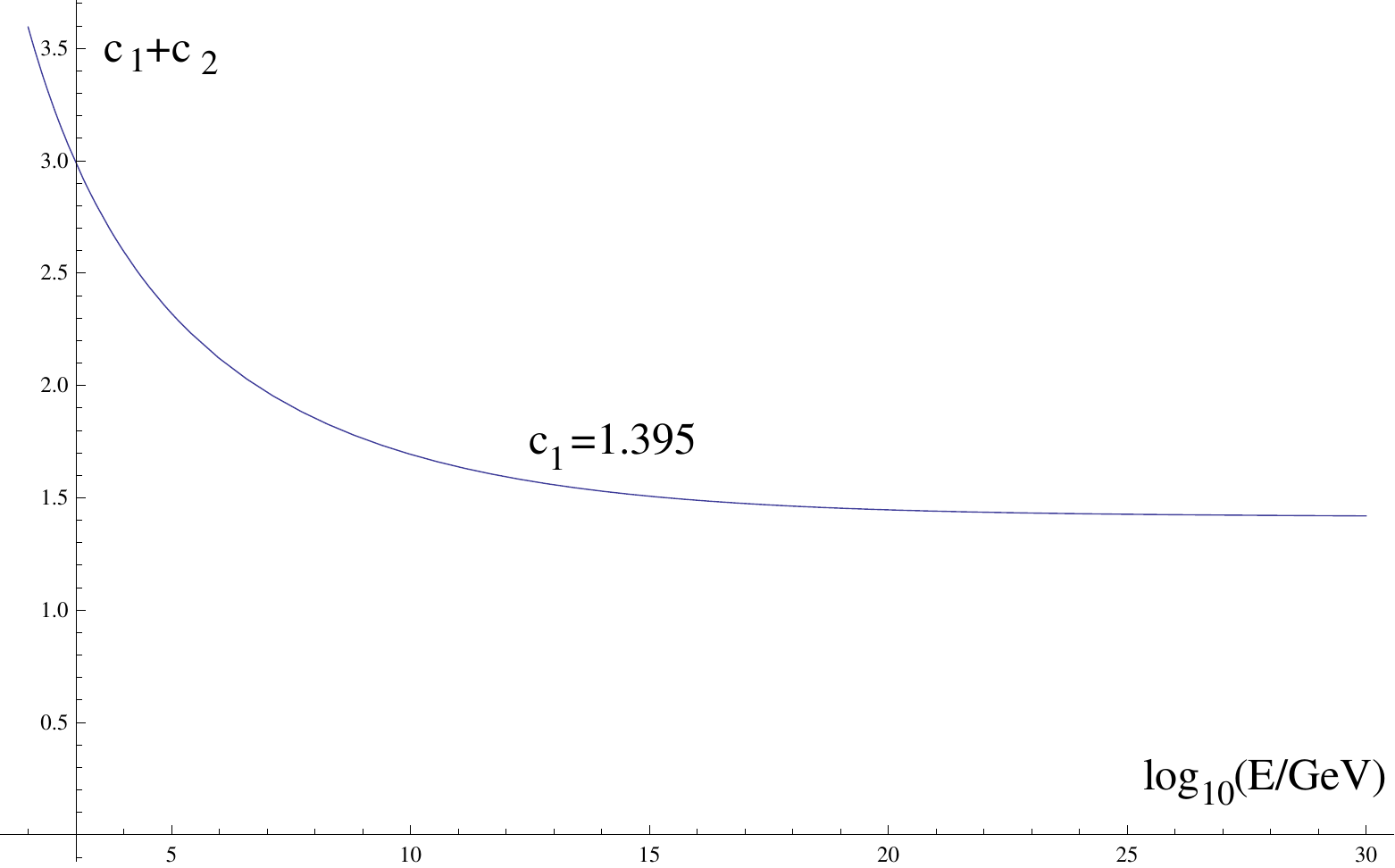}
\caption{Plotting of $\eta_\lambda$ as a function of $\log_{10}(E)$ for  $c_1=1.395$. The corresponding running (pole) Higgs mas is $\sim 163(155)$ GeV
(2-loop running).}
\label{fig:lambda_E_1}
\end{figure}

Going to 2-loop order, we should first determine the value(s) of the constant $c_1$ in Eq.\ref{eq:ansatz1}. In this order,
the procedure of keeping only the large terms in the high-energy regime, does not lead to an independent from $\ga_1$ value(s)
 $c_1$. Nevertheless, for a big range of $\ga_1$, $c_1$ varies by less than 5\% from the 1-loop case:
0.0448 - 0.0465 for the lower value and 1.342 - 1.395 for the higher one.
We solve now the 2-loop differential equation for $c_2$ using as an initial value of $\eta_\lambda$ at (very) high energies (i.e. low
value of $\eta_2$) the value $c_1=1.395$. The new value drives the ratio $\eta_\lambda$ to its constant value early on the energy scale
(see Fig.\ref{fig:lambda_E_1}). To be more specific, we see that $\eta_\lambda(M_{Planck})=1.459$ and it remains pretty stable for higher
energies. The Higgs running (pole) mass is $\sim 163(155)$ GeV. At the 2-loop level, the problem with the lower $c_1$ value persists:
$\eta_\lambda$ passes through negative values.

\noindent
\textbf{\textit{4. The MSSM case}}.
If we assume that the top and bottom Yukawa couplings are related, the reduction equation is
\[
\frac{d\ga_t}{d\ga_b}=\frac{\beta_t}{\beta_b}=
\frac{\ga_t\left(6\ga_t+\ga_b-c_i^{(t)}\ga_i\right)}{\ga_b\left(6\ga_b+\ga_t+\ga_\tau-c_i^{(b)}\ga_i\right)},
\]
where $c_i^{(t)}=(13/30,3/2,8/3)$ and $c_i^{(b)}=(7/30,3/2,8/3)$. Let us ignore, for simplicity, the contribution
of $\ga_\tau$ and the small difference between $c_1^{(t)}$ and $c_1^{(b)}$. Then,  it is straightforward to deduce that
if the ratio $\ga_t/\ga_b$ is constant, then this ratio is equal to the corresponding ratio of the $\beta$-functions
and is equal to 1.
\[
\frac{d}{dt}\left(\frac{\ga_t}{\ga_b}\right)=0\rightarrow
\frac{1}{\ga_b^2}\left(\ga_b\beta_t-\ga_t\beta_b\right)=0\rightarrow
\frac{\ga_t}{\ga_b}=\frac{\beta_t}{\beta_b}.
\]
This result combined with the previous equation leads to
\[
6\ga_t+\ga_b-c_i^{(t)}\ga_i=6\ga_b+\ga_t+\ga_\tau-c_i^{(b)}\ga_i\rightarrow
\ga_t=\ga_b.
\]
That is, if we start with equal $\ga_t$ and $\ga_b$ at an energy scale, equality will remain
 for all energies. Putting back the $\tau$ Yukawa coupling and the difference between the $c_1^{(t)}$
and $c_1^{(b)}$ constants, we expect a small deviation from that behavior.

Therefore, the procedure is the following: we start the running (with the SM RGEs) from the known
values of the top-, bottom- and tau-mass. At $M_{SUSY}$, we choose the appropriate $\tan\beta$ value
that keeps the ratio $\ga_t/\ga_b$ constant for all energies. Of course, we expect%
\footnote{
The fact that $\tan\beta$ could be predicted using reduction of couplings was suggested in
\cite{KMOZ} in a discussion with a different focus.}
this constant to be near 1.

In the MSSM scenario, at the scale $M_{SUSY}$, we have the relations
\begin{equation}
\label{eq:SUSY}
\begin{split}
\ga_t(SM)&=\ga_t(MSSM)\,\sin^2\beta\\
\ga_b(SM)&=\ga_b(MSSM)\,\cos^2\beta\\
\ga_\tau(SM)&=\ga_\tau(MSSM)\,\cos^2\beta.
\end{split}
\end{equation}
Above the $M_{SUSY}$ scale, the running
of all the parameters obeys the MSSM renormalization group equations, while below that scale, the SM regime is active.

\begin{figure}[!t]
\begin{tabular}{cc}
\includegraphics[scale=0.48,angle=0]{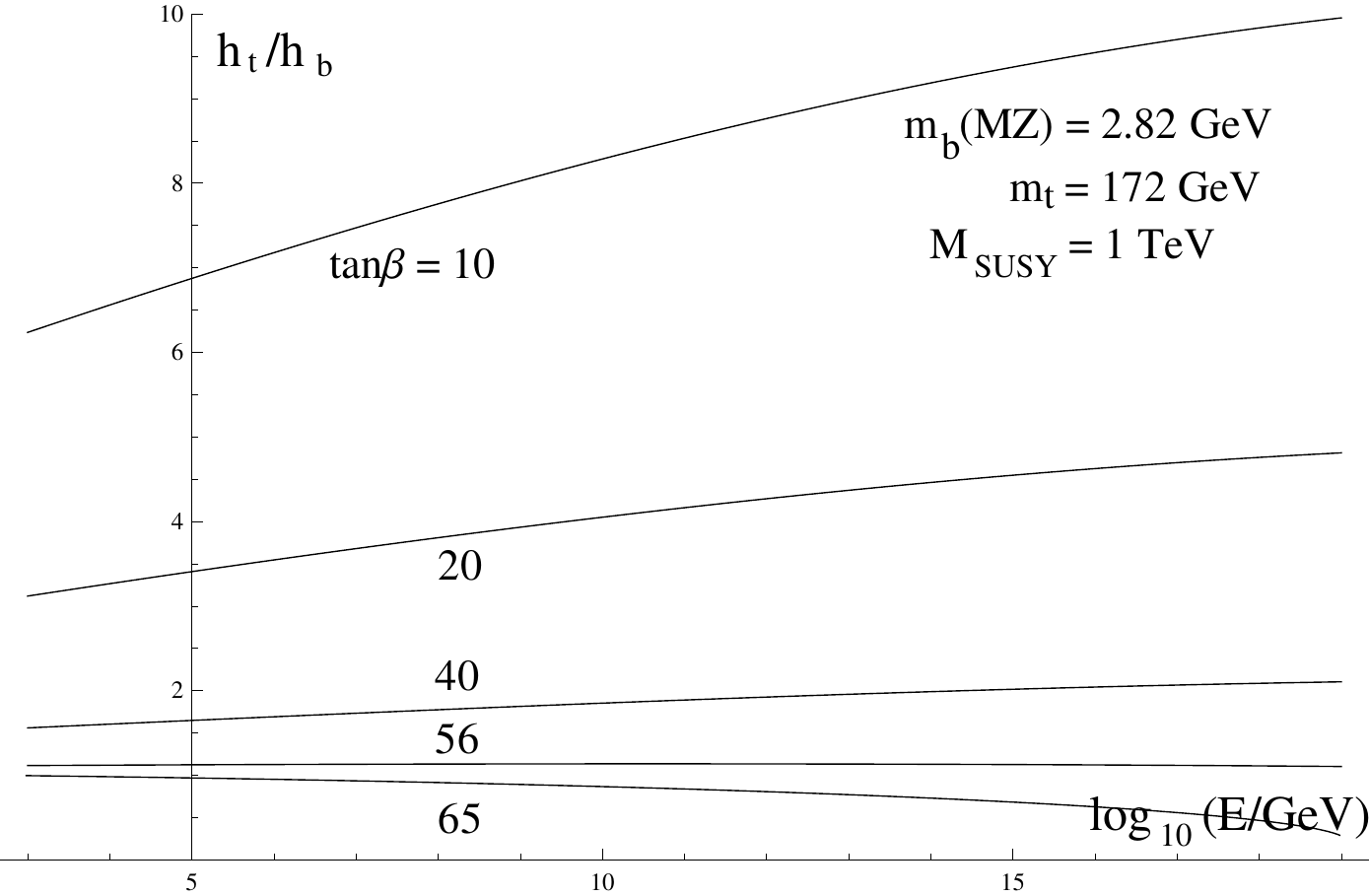}&
\includegraphics[scale=0.55,angle=0]{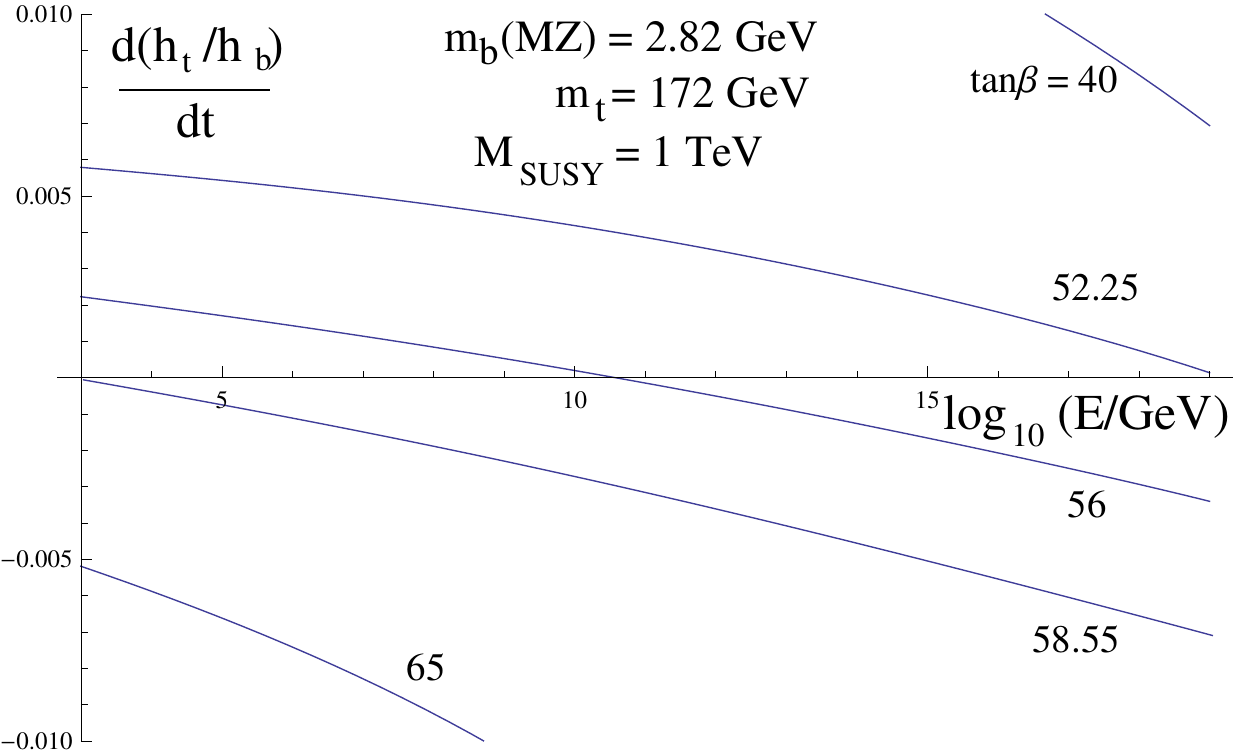}\\
(a)&(b)
\end{tabular}
\caption{(a) The ratio $h_t/h_b$ and (b) the derivative of the ratio as a function of energy for several values of $\tan\beta$ and
$M_{SUSY}=1$~TeV, $m_t=172$ GeV and $m_b(M_Z)=2.82$ GeV.}
\label{fig:htoverhb_1}
\end{figure}

In Fig.\ref{fig:htoverhb_1} we plot the ratio $h_t/h_b$ (a) and the derivative of the ratio (b)
as a function of  energy, for several values of $\tan\beta$ and $M_{SUSY}=1$~TeV, $m_t=172$ GeV
and $m_b(M_Z)=2.82$ GeV.
We clearly see that for the range $\tan\beta=52.25-58.55$, the derivative of the ratio stays almost zero
(actually less than $6\cdot 10^{-3}$).
The two values of $\tan\beta$: 52.25 and 58.55, are the limiting cases.
For values below the first one, the derivative stays positive, while above the second one the derivative stays negative
for the whole energy range.

\begin{figure}[!t]
\begin{tabular}{cc}
\includegraphics[scale=0.45,angle=0]{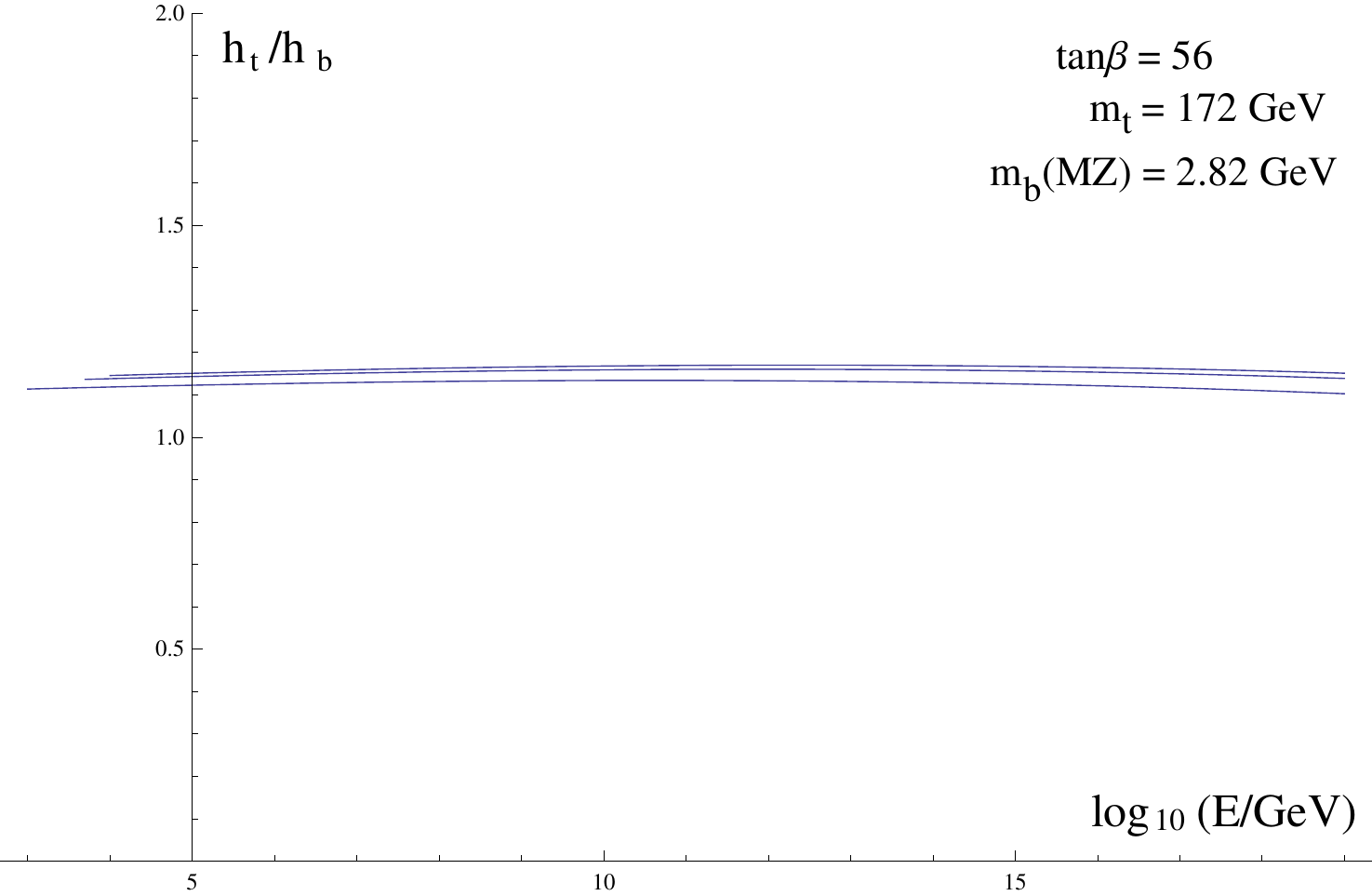}&
\includegraphics[scale=0.45,angle=0]{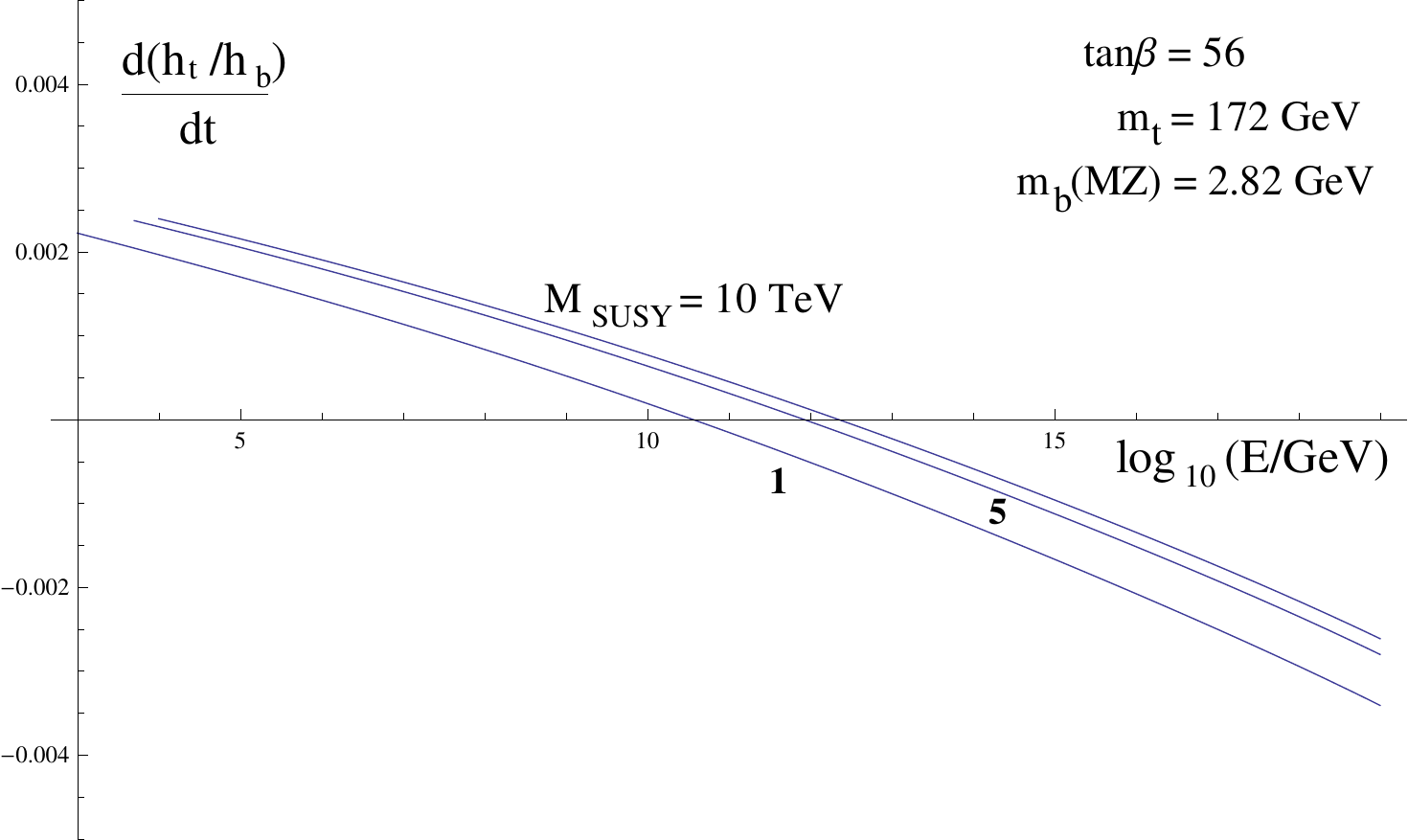}\\
(a)&(b)\\
\includegraphics[scale=0.45,angle=0]{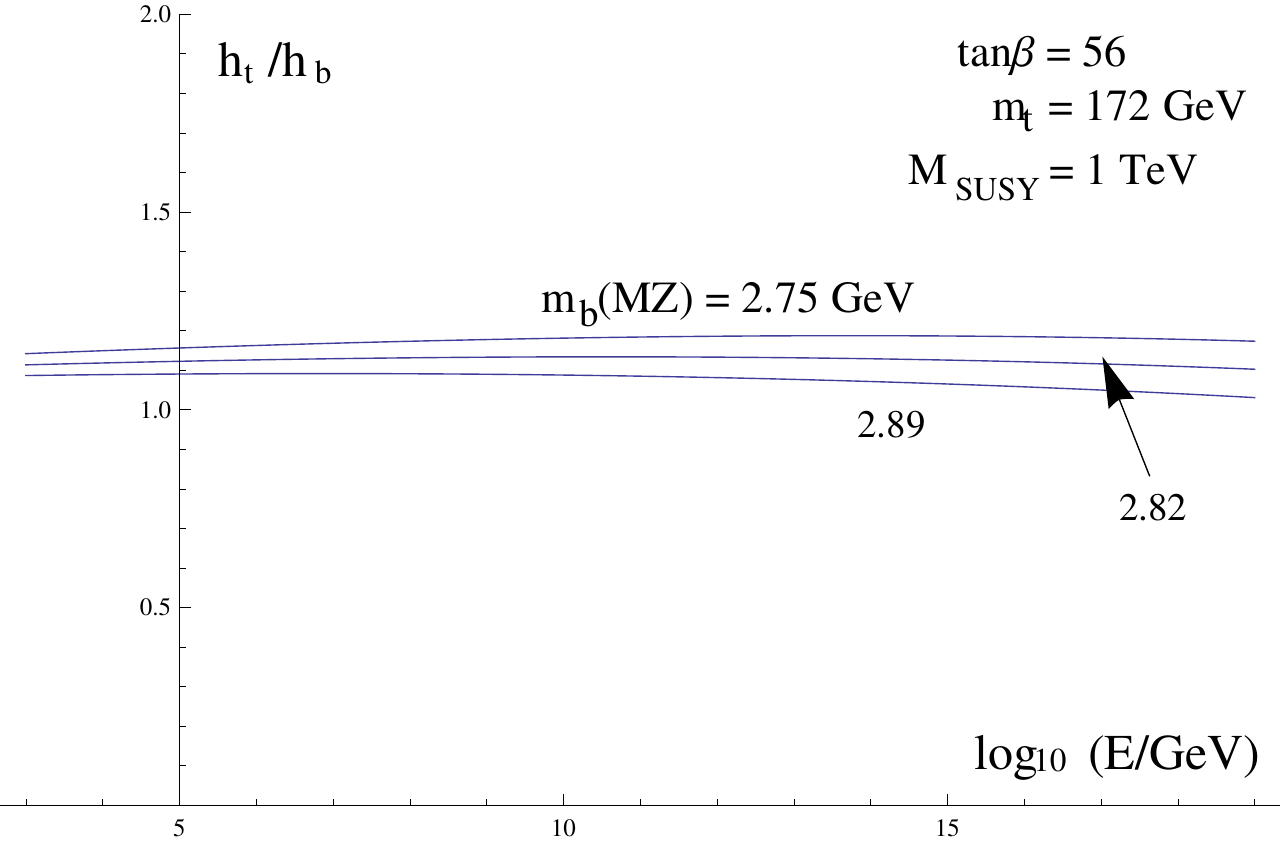}&
\includegraphics[scale=0.45,angle=0]{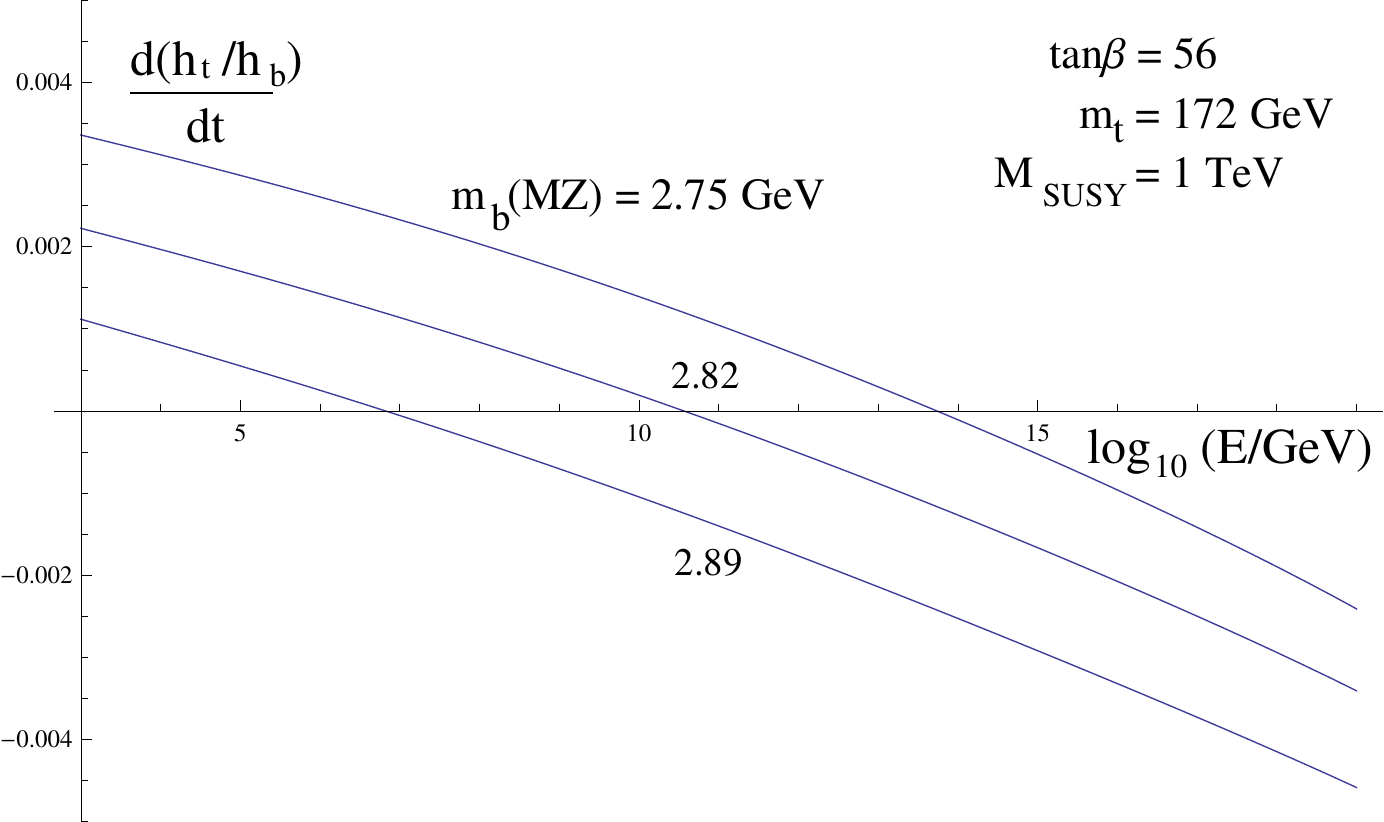}\\
(c)&(d)
\end{tabular}
\caption{Plots of the ratio $h_t/h_b$ ((a) and (c)) as well as the derivative of the ratio ((b) and (d)) as a function
of energy for $M_{SUSY}=1$, 5 and 10 TeV ((a) and (b)) and varying the bottom mass in the experimental error region ((c) and (d)).}
\label{fig:htoverhb_2}
\end{figure}

In Fig.\ref{fig:htoverhb_2} we plot the ratio $h_t/h_b$ (in (a) and (c)) as well as the
 derivative of the ratio (in (b) and (d)) as a function of  energy for the central value
of $\tan\beta=56$. In (a) and
(b) we show three curves corresponding to $M_{SUSY}=1$, 5 and 10 TeV, keeping the masses
of top and bottom at their central values. In (c) and (d) we vary the bottom mass
$m_b(M_Z)=2.75$, 2.82 and 2.89 GeV, keeping the top mass at its central value and $M_{SUSY}=1$ TeV.
The differences upon varying the top mass are negligible.

\begin{figure}[!t]
\centering
\includegraphics[scale=0.7,angle=0]{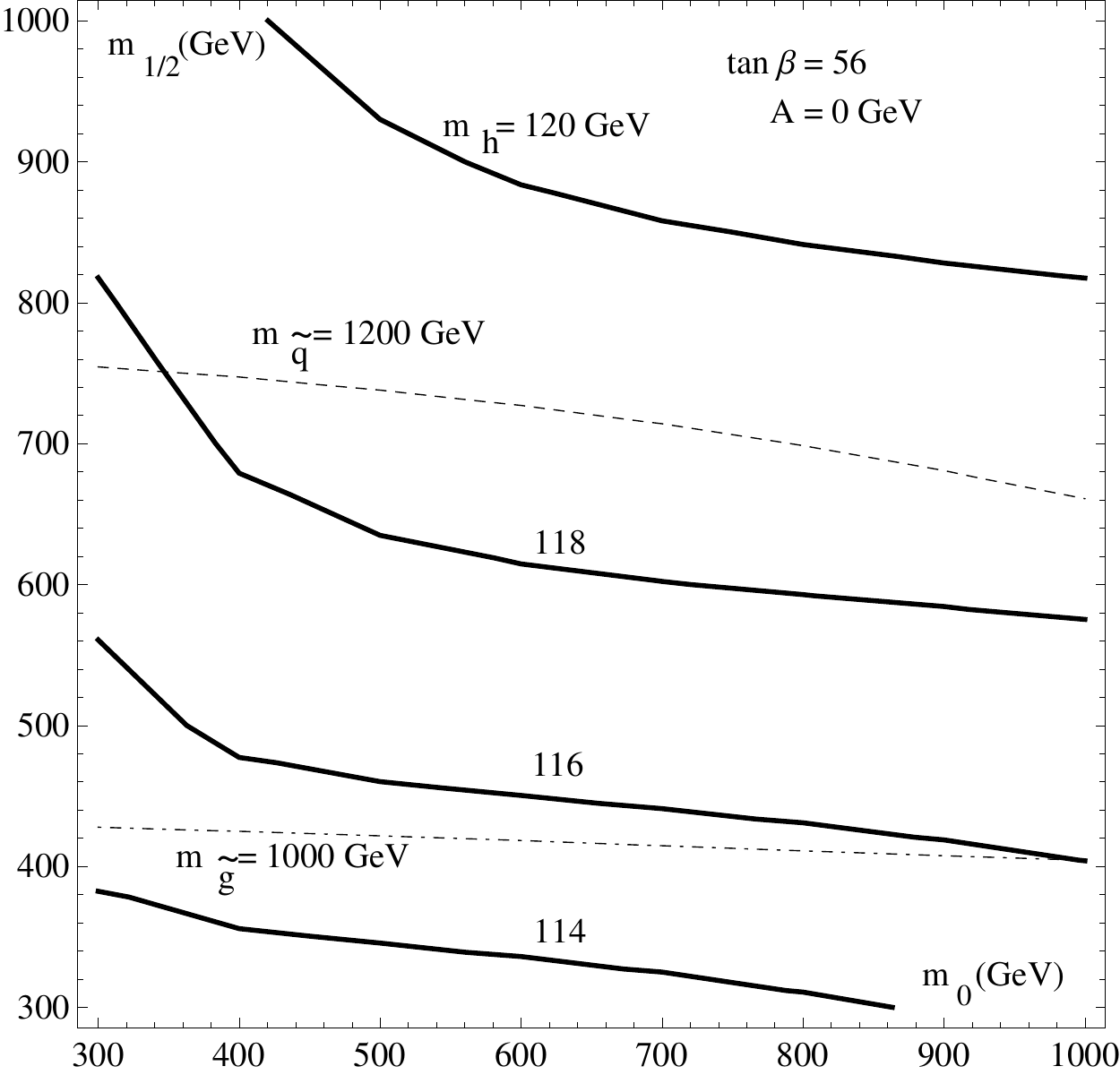}
\caption{Contours of constant $m_h$ (pole) mass in the plane of $(m_0,m_{1/2})$ for initial value $A=0$ GeV and for
$\tan\beta=56$. The dashed and the dotted-dashed contours correspond to (lightest) squark and gluino masses of 1.2 TeV and 1 TeV
correspondingly.}
\label{fig:m_h_SUSY}
\end{figure}

\begin{figure}[!t]
\centering
\includegraphics[scale=0.7,angle=0]{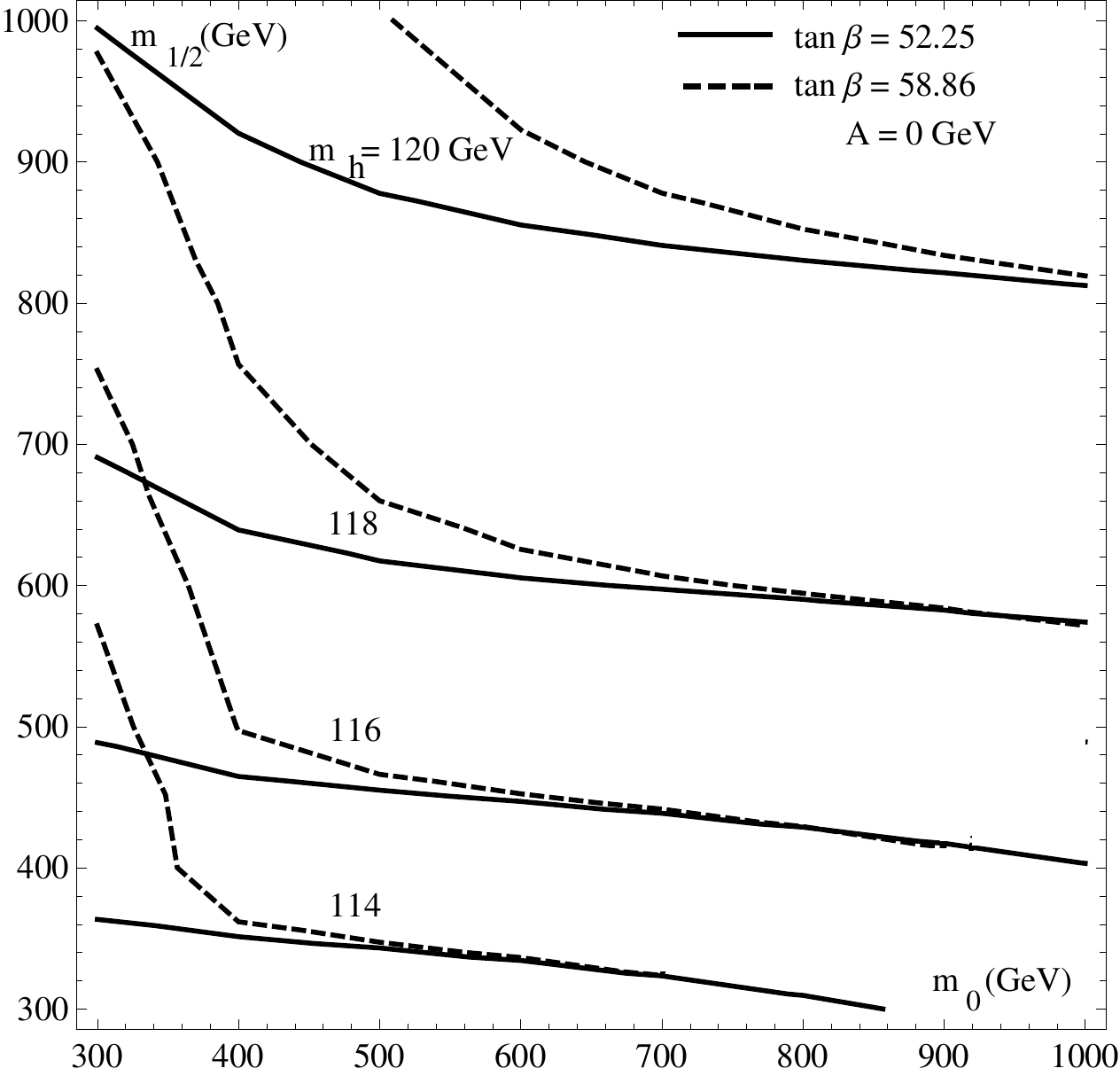}
\caption{The same as in Fig.\ref{fig:m_h_SUSY} for the two limiting $\tan\beta$ cases: 58.55 and 52.25}
\label{fig:m_h_SUSY2}
\end{figure}

Now, using the program SUSPECT \cite{SUSPECT}\footnote{%
We run the programm using the mSUGRA model, 2-loop running and evaluation of pole masses. In all cases $\textrm{sign}(\mu)=+1$.},
we can plot in the plane of $(m_0,m_{1/2})$ contours of constant
(pole) mass values for the lightest supersymmetric Higgs $m_h$ \footnote{%
We keep $m_H$ for the SM Higgs and denote by $m_h$ the lightest Higgs in the MSSM.}
for $\tan\beta=56$.
In Fig.\ref{fig:m_h_SUSY} we show these contours for $m_h=114,116,118,120$ GeV for initial $A=0$ GeV and $\tan\beta=56$.
The dotted-dashed contour corresponds to a gluino mass of 1 TeV, while the dashed contour to (the lightest) squark mass of 1.2 TeV.
According to recent data from ATLAS/LHC and CMS/LHC\cite{ATLAS_and_CMS}, the two values represent the lower bounds for detection of the corresponding
particle. Finally, in Fig.\ref{fig:m_h_SUSY2} we plot the same contours for the two limiting $\tan\beta$ cases: 58.55 and 52.25.

\noindent
\textit{\textbf{5. Conclusions.}}
 The idea of couplings reduction  in a theory is  very  appealing since it increases its predictive power.
 Successful reduction led to all-loop finite theories and a prediction of the top-quark mass. The latter property
 was used as a selection criterion for a successful GUT. In the present work, we have studied the reduction of certain
 couplings within the SM, and  have obtained a  prediction for the Higgs particle mass. Previous studies either overlooked this possibility, or did not include the heavy top-quark contribution. We have also started an analogous
 analysis in the MSSM, which we plan to extend in a forthcoming publication.

\vspace*{1cm}
\noindent
\textit{\textbf{Acknowledgements.}}
It is a pleasure to thank A. Djouadi, W. Hollik, S. Heinemeyer, L. Fayard, J. Kubo, E. Ma, M. Mondragon and W. Zimmermann for very interesting discussions.
G.Z. is grateful to the Sommerfeld-LMU and MPI Munich the warm hospitality. This work was partially supported by the NTUA's basic research
support program ``PEVE'' 2009 and 2010 and the European Union ITN programme ``UNILHC'' PITN-GA-2009-237920.

\end{document}